# Simulation of beam gas coulomb scattering in HALS *

YU Lu-Xin(于路新)[1)] GAO Wei-Wei(高巍巍)[2)] WANG Lin(王琳)[3)] LI Wei-Min(李为民)

NSRL, School of Nuclear Science and Technology,
University of Science and Technology of China, Hefei 230029, P. R. China

**Abstract:** In conventional research on the beam gas coulomb scattering (BGCS), only the related beam lifetime using the analytical method is studied. In this paper, using the PIC-MCC method, we not only simulated the beam lifetime but also explored the effect of BGCS on the beam distribution. In order to better estimate the effect on particle distribution, we study the ultra-low emittance electron beam, here we choose the HeFei Advanced Light Source (HLAS). By counting the lost particles in a certain time, the corresponding beam lifetime we simulated is 5.1078h/14.5507h in x/y, which is very close to the theoretic value (5.0555h /13.7024h in x/y). By counting the lost particles relative to the collided particles, the simulated value of the loss probability of collided particles is 1.3228e-04, which is also very close to the theoretical value (1.3824e-04). Besides, the simulation shows there is a tail in the transverse distribution due to the BGCS. The close match of the simulation with theoretic value in beam lifetime and loss probability indicates our simulation is reliable.

## 1 Introduction

In e-storage ring, there is always some residual gas near the beam orbit, no matter how well vacuum condition is. The electrons will inevitably collide with them. The interaction between electrons and residual gas is quite complex. There are mainly three mechanisms: the coulomb scattering，bremsstrahlung scattering，and the inelastic scattering with outer-shell electron. These collisions will cause the change of the beam electron motion state, if the change is enough large, the corresponding electron will loss. In the four types of beam gas interaction, the most common and basic is the coulomb scattering on the nucleus which is called as BGCS. Here, what we are studying is the BGCS [1] .

Conventionally, the numerical analysis is the most common way to study BGCS [2] [3]. And the analysis is limited to obtain the calculation formula of beam lifetime. In recent years minority or individual researches try to simulate beam lifetime using MC method [4].

But, does the beam gas scattering just affect the beam lifetime? What about particle distribution? Analyse the process of particle loss. First the momentum of the collided particle is changed due to the scattering. Then along with the beam transport, the position of the particle is changed. At last, the particle will be lost if it exceeds the limited conditions. From this process, it can be seen that the beam lifetime is also based on the change of particle distribution. Unfortunately, there is no research on the influence that BGCS does to the particle distribution.

In this paper, using the PIC-MCC method and choosing appropriate cross section and scattering angle, we not only simulated the beam lifetime but also explored

* Supported by the Natural Science Foundation of China ( 10979045 and 11175182 ) and the Graduate Innovation Foundation of USTC

1)E-mail: yuluxin@mail.ustc.edu.cn;

2)E-mail: gaomqr@mail.ustc.edu.cn;

3)E-mail: wanglin@ustc.edu.cn

the effect of BGCS on the beam distribution. Moreover, previous study on BGCS just gave a kick to transverse oscillation. In order to get the post-momentum of particle more accurately, we deal with the collision process in laboratory coordinate system.

This paper is organized as follows: In Section 2 the basic theory is given. In Section 3 the simulation technique is described. The results of the simulation are compared with the theoretic study in section 4. And Section 5 is devoted to a discussion and conclusion.

## 2 Basic Theory of the simulation

### 2.1 The PIC-MCC method [5][6]

The particle-in-cell merged with Monte Carlo collision (PIC-MCC) is a widely used numerical technique in many-charged-particles simulation for decades. It approaches to balance the PIC and MCC. With PIC-MCC method to track the motion of massive particles, all the micro information of the particle system is included. And any information whether microcosmic or macroscopic, in principle, can be obtained.

Using the PIC-MCC to describe the collision of particles, the collision probability during one time-step $\Delta t$ can be got by the following expression [6]:

$$p(t) = 1 - \exp(-\sigma \rho v \Delta t)\,. \qquad (10)$$

Where $\sigma$ is the collision cross section, $\rho$ is the volume density of target particles, $\upsilon$ is the relative velocity of collided particles.

Generate a uniformly distributed random number $R_i \in [0\ 1]$, and by comparing the *p(t)* with $R_i$ to determine if the particle can collide. If $R_i$ is less than *p(t)*, the particle suffered a collision, then was processed by MCC method. If $R_i$ is larger than *p(t)*, the particle didn't collide, then was processed by PIC method.

### 2.2 The derivation of collision formulas

By transferring the collision into the center of mass reference frame (CM), W.G.Vincenti and G.H.Kruger give the formulas of collision between two non-relativistic particles [7]. The main points are as follows.

In CM frame, the initial momenta of the two collided particles are equal and opposite. Furthermore, the force on each of the two particles is equal in magnitude but opposite in direction. As a result the final momenta deflecting at an angle $\eta$ are also equal and opposite. With these results conservation of total momentum and energy have been satisfied, and the changes in momenta can be written in terms of the initial momenta and $\eta$. The $\eta$ is defined in CM frame.

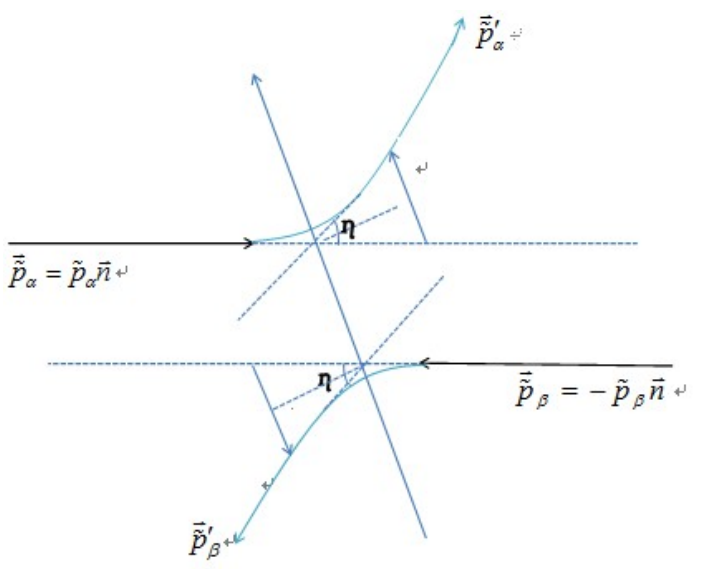

Fig.2 The trajectories of two particles during a collision in CM frame

But it will be more complex to get the post-momenta of relativistic particles, which need twice Lorenz transformation [8]. And in our simulation, we just need the electron's post-momentum. Combining the above points, we deal with the collision in laboratory coordinate system. Below is the description.

For electron–atom elastic collision, the atom is considered to have a so large mass relative to electron that the electron only scatters in angle $\chi$ with no loss of energy [5]. So the electron's trajectory in laboratory coordinate system can be described as shown in Fig.3.

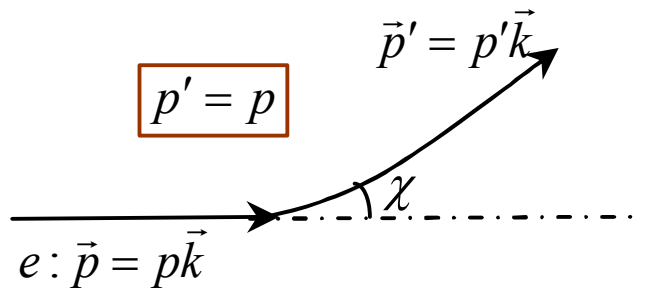

Fig.3 The electron's deflection in Lab

Where, $\vec{k}$ is the unit vector of $\vec{p}$, while $\vec{k}'$ is the unit vector of $\vec{p}'$, so the electron's collision process can be described as $|\vec{p}| = |\vec{p}'|$ with direction changed from $\vec{k}$ to $\vec{k}'$.

Since the $p'=p$, if the component of $\vec{k}'$ is known, $\vec{p}'$ will be easy to get. Now what we should do is to find the relationship between $\vec{k}'$ and $\vec{k}$ . The Fig.4 shows the relationship between $\vec{k}'$ and $\vec{k}$ .

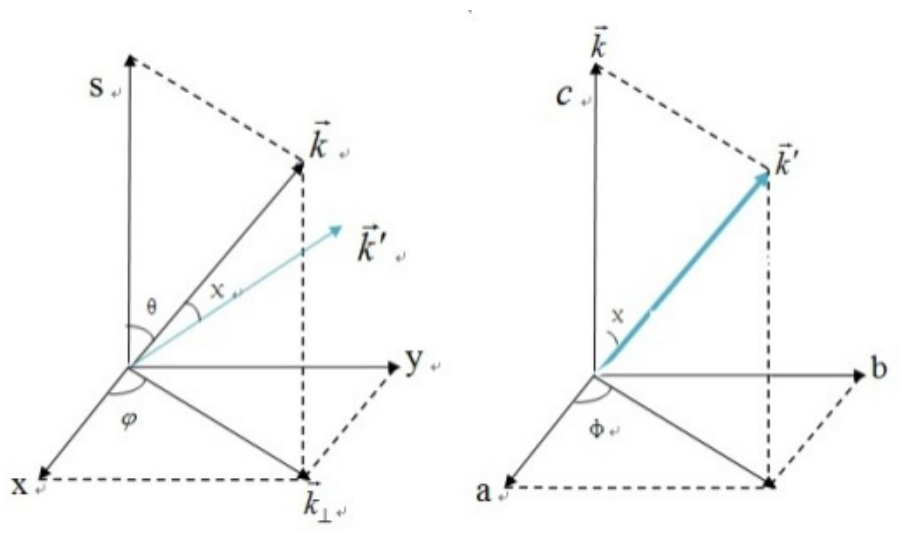

Fig.4 The vector diagram of $\vec{k}$ and $\vec{k}'$

On the left of Fig.4, the components of $\vec{k}$ in (*x, y, s*) can be got by Eq. (3) [7][9]

[9] Takizuka and H. Abe, "A Binary Collision Model for Plasma Simulation with a Particle Code", J.Computational Physics 25 (1977) 205.

$$\begin{pmatrix} \cos\theta\cos\varphi & \cos\theta\sin\varphi & -\sin\varphi \\ -\sin\varphi & \cos\varphi & 0 \\ \sin\theta\cos\varphi & \sin\theta\sin\varphi & \cos\theta \end{pmatrix}\begin{pmatrix} k_x \\ k_y \\ k_s \end{pmatrix}=\begin{pmatrix} 0 \\ 0 \\ k \end{pmatrix}\Leftrightarrow\begin{pmatrix} k_x \\ k_y \\ k_s \end{pmatrix}=\begin{pmatrix} k\sin\theta\cos\varphi \\ k\sin\theta\sin\varphi \\ k\cos\theta \end{pmatrix}. \quad (3)$$

Similarly, on the right of Fig.2 there is a relationship for $\vec{k}'$ in (*a, b, c*).

$$\begin{pmatrix} \cos\chi\cos\phi & \cos\chi\sin\phi & -\sin\phi \\ -\sin\phi & \cos\phi & 0 \\ \sin\chi\cos\phi & \sin\chi\sin\phi & \cos\chi \end{pmatrix}\begin{pmatrix} k'_a \\ k'_b \\ k'_c \end{pmatrix}=\begin{pmatrix} 0 \\ 0 \\ k' \end{pmatrix}\Leftrightarrow\begin{pmatrix} k'_a \\ k'_b \\ k'_c \end{pmatrix}=\begin{pmatrix} k'\sin\chi\cos\phi \\ k'\sin\chi\sin\phi \\ k'\cos\chi \end{pmatrix}. \quad (4)$$

Where the angle Φ takes on values from 0 to 2π.

So $\vec{k}'$ in (*x, y, s*) has following relationship.

$$\begin{pmatrix} \cos\theta\cos\varphi & \cos\theta\sin\varphi & -\sin\varphi \\ -\sin\varphi & \cos\varphi & 0 \\ \sin\theta\cos\varphi & \sin\theta\sin\varphi & \cos\theta \end{pmatrix}\begin{pmatrix} k'_x \\ k'_y \\ k'_s \end{pmatrix}=\begin{pmatrix} k'\sin\chi\cos\phi \\ k'\sin\chi\sin\phi \\ k'\cos\chi \end{pmatrix}. \quad (5)$$

It can be driven

$$\begin{pmatrix} k'_x \\ k'_y \\ k'_s \end{pmatrix}=\begin{pmatrix} k'\cos\chi\cos\varphi\sin\theta+k'\sin\chi\cos\theta\cos\varphi\cos\phi\text{-}k'\sin\chi\sin\varphi\sin\phi \\ k'\cos\chi\sin\varphi\sin\theta+k'\sin\chi\cos\theta\sin\varphi\cos\phi+k'\sin\chi\cos\varphi\sin\phi \\ k'\cos\theta\cos\chi\text{-}k'\sin\chi\cos\phi\sin\theta \end{pmatrix}. \quad (6)$$

Taking $k'=k=1$ and Eq. (3) into Eq. (6).

$$\begin{pmatrix} k'_x \\ k'_y \\ k'_s \end{pmatrix}=\begin{pmatrix} k_x\cos\chi+k_xk_s/k_\perp\sin\chi\cos\phi-k_y/k_\perp\sin\chi\sin\phi \\ k_y\cos\chi+k_yk_s/k_\perp\sin\chi\cos\phi+k_x/k_\perp\sin\chi\sin\phi \\ k_s\cos\chi\text{-}k_\perp\sin\chi\cos\phi \end{pmatrix}. \quad (7)$$

Taking $k_x=p_x/p$, $k_y=p_y/p$, $k_s=p_s/p$, $k_\perp=p_\perp/p$ into Eq. (7). Then the post-collision momentum in laboratory coordinate system for an elastic collision is given by

$$\vec{p}'=p'\vec{k}'=\vec{p}\cos\chi+\vec{h}\sin\chi\,. \quad (8)$$

Where $\vec{h}=(h_x,h_y,h_s)$ with

$$\begin{aligned} h_x &= (p_xp_s\cos\phi-p_yp\sin\phi)/p_\perp \\ h_y &= (p_yp_s\cos\phi+p_xp\sin\phi)/p_\perp \\ h_s &= -p_\perp\cos\phi \end{aligned}\quad. \quad (9)$$

Where Φ is uniformly distributed in [0, 2π]. χ is an angle defined in laboratory coordinate system.

It should be clearly noted that although the Eq. (8) is obtained under laboratory coordinate system, but the collision is instantaneous, it also applies to the Frenet-Serret coordinate system.

With the collision formulas, we will investigate the beam lifetime and the particle distribution's change caused by the coulomb scattering with the residual gas atoms.

## 3 Simulation method

### 3.1 The description of the simulation method [4]

The simulation is based on the macro-particle method. The macro-particles transfer through the ring and encounter coulomb scattering with residue gas atoms simultaneously.

The initial 6-D coordinates of n macro-particles are given randomly with specified variances. Each macro-particle (*i*) has a particle number ($N_i$). $\Sigma N_i$ is the total number of particles in a bunch.

We set one fixed interaction point (IP) in the ring, use one turn transfer map to transmit the particle. We define *p* is the probability that an electron scattering with gas nucleus in one turn. So the probability that each macro-particle undergoes a random process is $N_ip$.

For each macro-particle, a random number $R\in\infty[0,1]$ is chosen each turn, if $R< N_ip$, we separate one electron from the *i*-th macro-particle as a new macro-particle. The new macro-particle has the same coordinates to the parental macro-particle.

During the collision, the ($p_x$, $p_y$, $p_s$) of the new macro-particle is changed, while the (*x*, *y*, *s*) is considered the same as before. And the new macro-particle will not undergo the BGCS again, just propagates in the ring.

As a result of the specific aperture，this process will lead to the loss of particles. Then the beam lifetime of the beam will be estimated in the simulation by counting the number of the particles extending beyond the specific aperture.

### 3.2 The choice of cross section and scattering angle

In previous simulation, only the scattering that would cause the particle loss is considered to obtain the beam lifetime. Whether the particle would lose depends on the relationship of the scattering angle and the critical angle. The critical angle is the minimum angle which would cause particle loss. And the electron loss cross section is used to calculate the loss probability. Now that we are going to explore the effect of gas scattering on particle distribution, we should take into account all the scattering. So we choose the appropriate cross section and scattering angle, rather than the particle loss section and critical scattering angle, to simulate the BGCS.

For the relativistic electrons collisions, the knowledge of the screening of the coulomb potential of the nuclei by the atomic electrons is important, since they are always scattered into a very small angle. The differential cross section that contains the shielding effect of the electron cloud can be determined by [9] [10].

$$\frac{d\sigma}{d\Omega}=\frac{4Z^2}{a_H^2}\frac{\gamma^2}{\left[(\frac{4\pi}{\lambda_e}\sin\frac{\chi}{2})^2+\frac{1}{R^2}\right]^2} \quad . \tag{11}$$

where $\Omega$ is the solid angle, $\chi$ the scattering angle, $Z$ the atomic number, $\lambda_e$ the wave-length of the electron, $\gamma$ the Lorentz factor and the radius of atom $R=a_h{*}Z^{-1/3}$. $a_h$ is the Bohr's radius of the atoms.

By integrating the Eq. (11) over the whole $\Omega$, we can obtain the total cross section

$$\sigma=(\frac{4Z\gamma}{a_h})^2(\frac{\lambda_e}{4\pi})^4(\frac{1}{2\varepsilon}-\frac{1}{2+2\varepsilon}) \quad . \tag{12}$$

With $\varepsilon=(\lambda_e/4\pi R)^2$ is the shielding parameter.

During the simulation, a random number generator is used at each turn to decide whether the scattering occurs. If yes, the scattering angle is defined according to the following formula [8].

$$\chi=\frac{\pi}{2}-a\sin\left[1-\frac{2R_1}{1-(R_1-1)(4\pi R/\lambda_e)^2}\right]. \tag{13}$$

Where $R_1$ is uniformly distributed from 0 to 1.

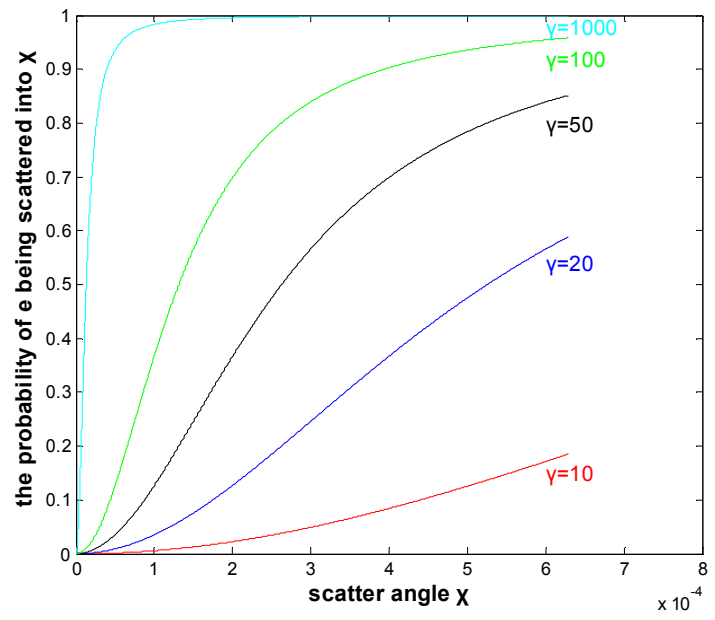

Fig.5 The collision probability of the electron with different energy being scattered into an angle χ.

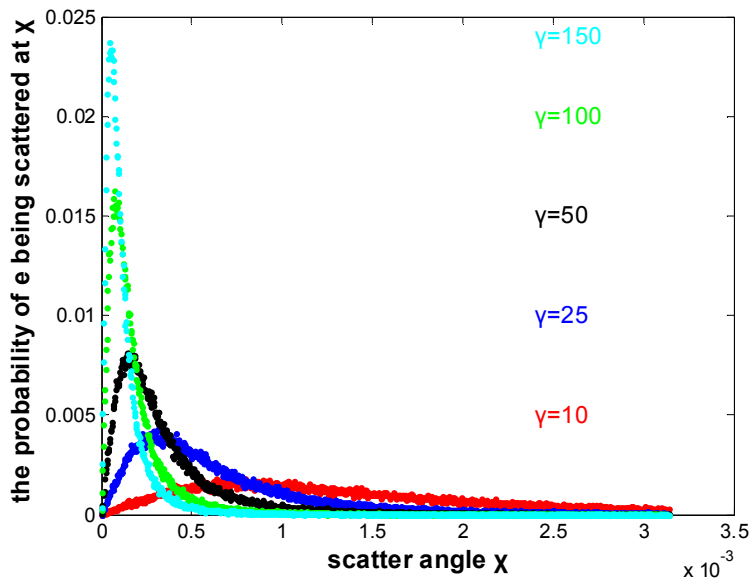

Fig.6 The distribution of scattering angle under different energy

The distribution of scattering angle under different energy is shown in Fig.5 and Fig.6. It can be seen that the higher energy the electron is, the smaller angle the electron will be scattered at. Conversely, the scattering angle tends to be uniform from 0 to pi for nonrelativistic electron. This is consistent with theoretical analysis, which also shows the formula is reasonable and reliable. For the HALS, the energy of the electron is 1.8GeV, and the probability that the scattering angle smaller than 2.5808e-5rad is 98%.

## 4 Simulation results

In the simulation, because the newly generated macro-particles hardly occurs secondary collision, we assume the new micro-particles won't undergo the BGCS again. For this reason, the number of initial micro-particles shouldn't be less than the number of collided electrons in one turn. For HALS, about 7810 ($n_e$*p) electrons occur collision in a turn, we set 8e5 (about 100 times the collided particles) initial macro-particles. The total number of particles is 1e10, so each initial macro-particle contains 12500 electrons.

The vacuum is expressed by $N_2$-equivalent gas pressure. In order to verify the program, it is supposed that there is only $N_2$ in the ring. Each $N_2$ molecule has two atoms with atomic number 7. Table 1 shows the simulation parameters of HALS.

Table 1. Simulation parameters of HALS

| Parameter | Description | Value |
|---|---|---|
| E | Beam Energy | 1.8GeV |
| C | Circumference | 486m |
| $n_e$ | particles in bunch | $10^{10}$ |
| H | Harmonic number | 810 |
| $\beta_{ip}$ | Twiss parameter beta | 9.1307/3.4663 |
| $\alpha_{ip}$ | Twiss parameter alpha | -1.83e-3/5.6151e-5 |
| $\gamma_{ip}$ | Twiss parameter gamma | 0.1095/0.2885 |
| $\varepsilon_x$ | Hor. emittance | $6.2e^{-11}$m·rad |
| $\varepsilon_y$ | Ver. emittance | $6.2e^{-13}$m·rad |
| d | Dynamics aperture | 24/15mm |
| P | $N_2$-equivalent gas pressure | 1ntorr |

Moreover, in order to eliminate the MC fluctuation, we run five times, and take the statistical average.

### 4.1 The Lifetime of Beam

Once a particle's amplitude exceeds an aperture which is the dynamics aperture in this paper, this particle will be lost.

In the simulation, lost particles are counted each turn. Fig.7 shows the number of lost particles growing with time. The blue line represents the simulation values, while the red line represents the linear fitting values.

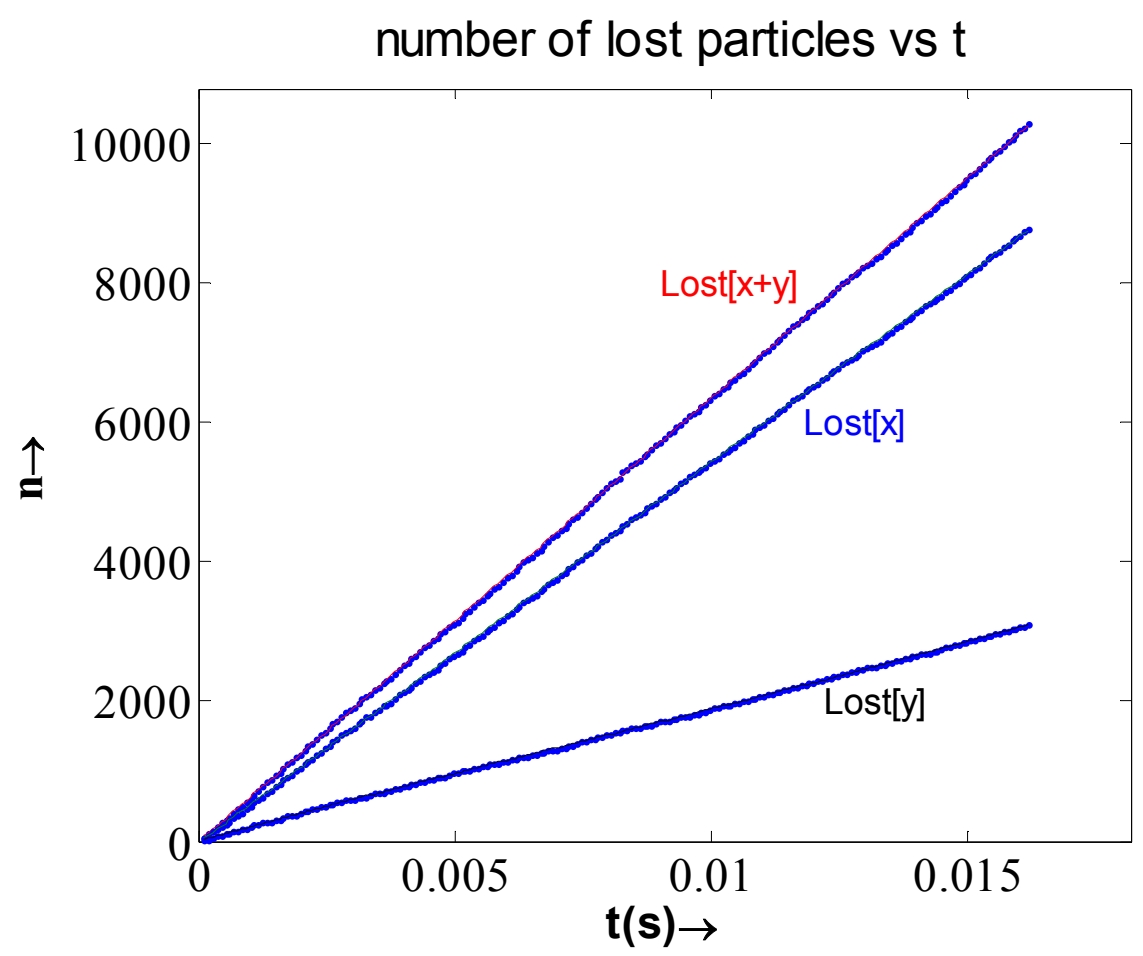

Fig.7 the number of lost particles with time

The simulation lifetime of beam can be calculated by

$$n = n_e e^{-t/\tau} \tag{14}$$

Where $n$ is the number of total particles after time t, while $n_e$ is the number of initial particles. With the date in Fig.7, the simulation beam lifetime is 5.1078h/14.5507h in x/y. Note that the lifetime in our simulation is the lifetime at IP.

Now, what we have to do is to verify whether this simulation result is reasonable. The beam coulomb lifetime related to the vacuum can be got as below [11].

$$\tau_{elstic,x,y} = \frac{2\pi r_e^2}{\gamma^2 kT} \frac{1}{A_{x,y}} \sum_{atom,j} (Z_j^2 \sum_{gas,i} \alpha_{ij} \langle \beta_{x,y} P_i \rangle) \tag{15}$$

Taking $\beta = \beta_{ip}$ , $A = d^2 / \beta_{ip}$ , then the corresponding lifetime at IP is 5.1875h/13.8751h in x/y.

It can be seen the simulation lifetime (5.1875h/13.8751h) is very close to the theoretical value (5.0555h /13.7024h).

On the other hand, the critical scattering angle can be computed by below [1].

$$\theta_{cx,y} = d_{x,y} / \beta_{x,y} \tag{16}$$

Define $\sigma_{loss}$ as the beam loss cross section. $\sigma_{loss}$ equals to the integration of Eq. (11) from $\theta_c$ to π.

Taking the geometric mean of $\theta_c$ in $x$ and $y$, we have $\theta_c$=3.3726e-04rad, then $\sigma_{loss}$=3.4710e-27m$^2$. And with $\sigma_{loss}/\sigma$, the theoretical loss probability of collided particles is 1.3824e-04.

In the simulation, take 10000 turns as example, the number of electrons that collided with the gas nucleus is 77794186, the number of lost electrons is 10291, so the simulation value of the loss probability of collided particle is 1.3228e-04, which is very close to the theoretical value (1.3824e-04).

**4.2 The Distributions of Particles**

In Fig.8, the red dots represent the initial distribution of particles, while the blue dots are the particle distribution after 10000 turns. Visually, the particles have a diffusion in both $x$ and $y$ due to the beam gas coulomb scattering. And one thing is for sure, the peripheral particles in the bunch after 10000 turns are mostly new macro-particles.

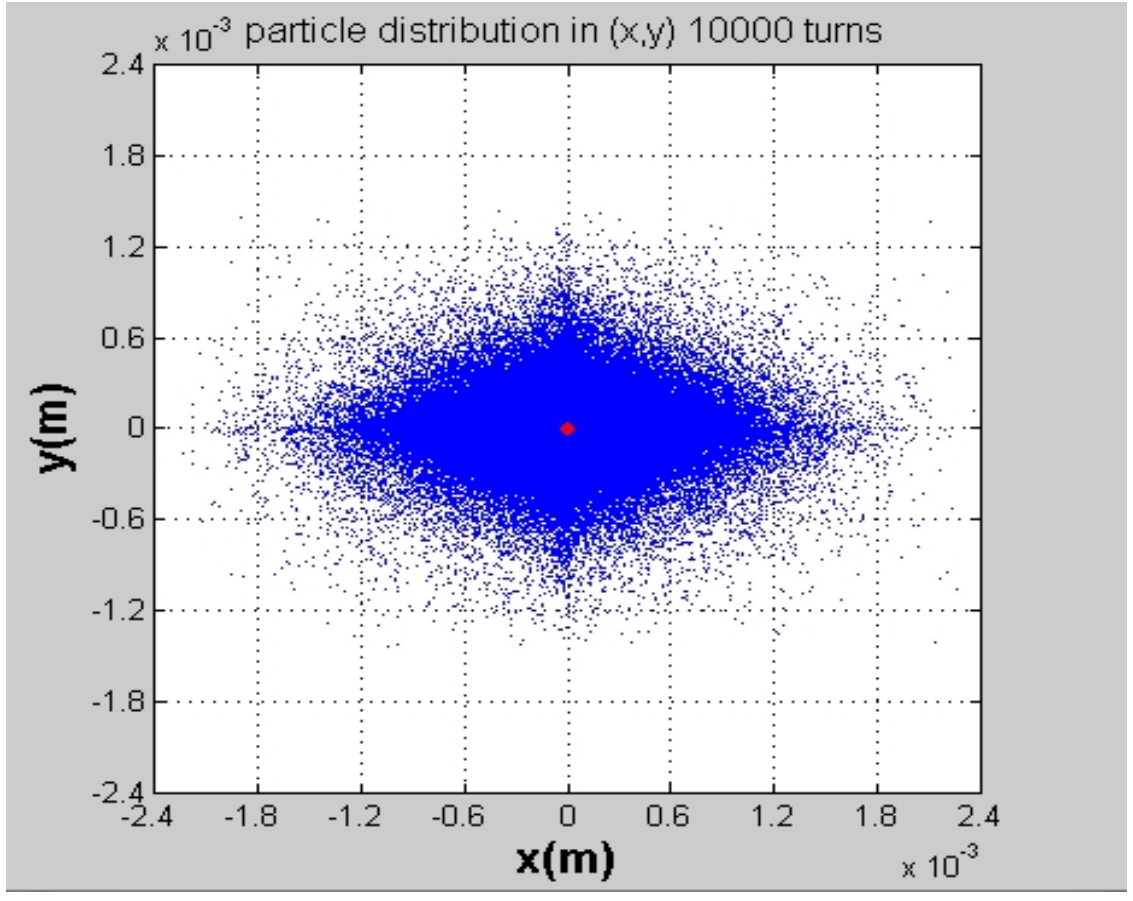

Fig.8 the transverse distribution of particles

Fig.9 shows the horizontal and vertical statistical distributions of the particles in a bunch. The red dots present the initial distribution, and the blue ones is the distribution after tracking. It can be seen there is a tail because of the beam gas scattering. The tail accounts for a little proportion of the total particles. In horizontal direction, the proportion that ρ is less than -15 is 1%. In vertical direction, the proportion that ρ is less than -12 is 2.6%. The result shows the tail in y is larger than x. In our simulation, the bunch is much more concentrated in y, and there is no coupling in x and y. So it comes to a conclusion that the effect of BGCS on particle distribution depends on how much the bunch is diverging. Which means the smaller the emittance is, the more significant is the effect of BGCS on the particle distribution.

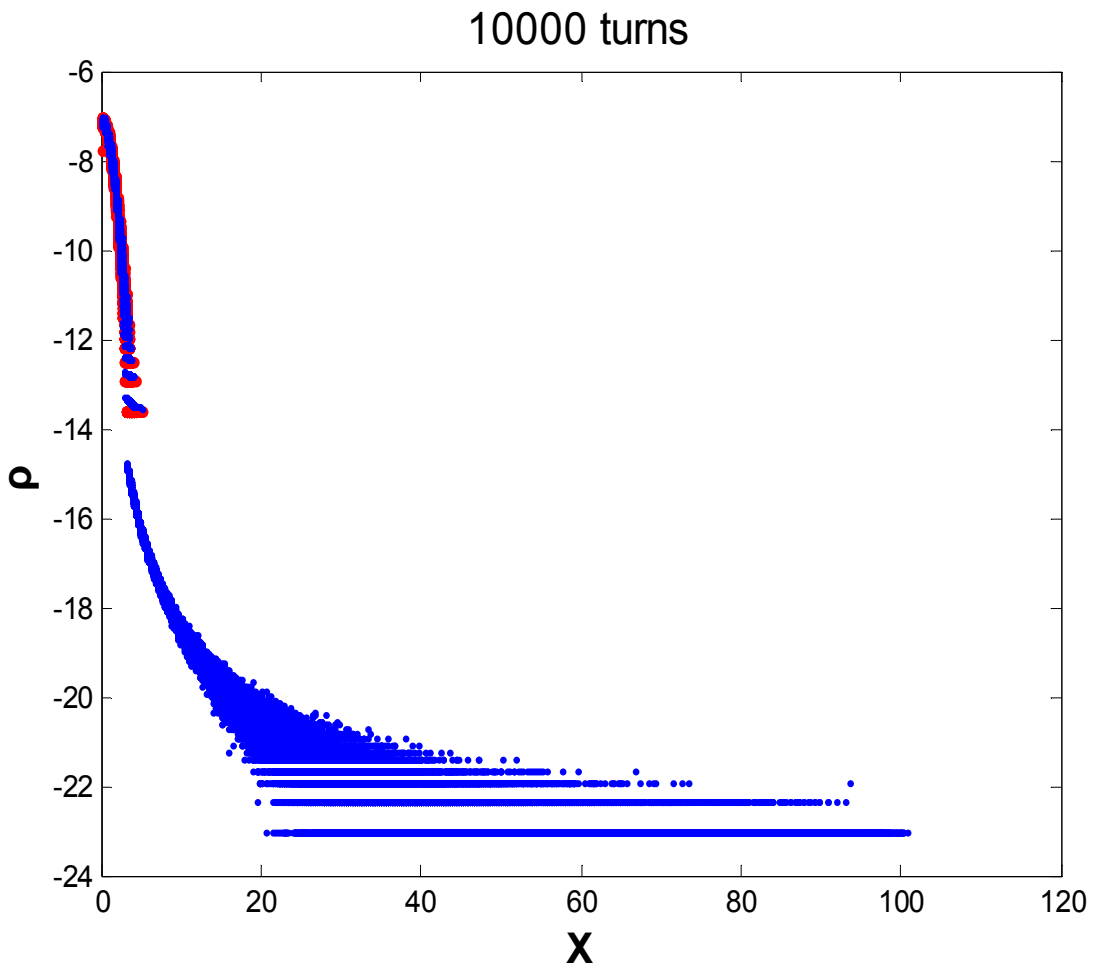

(a)

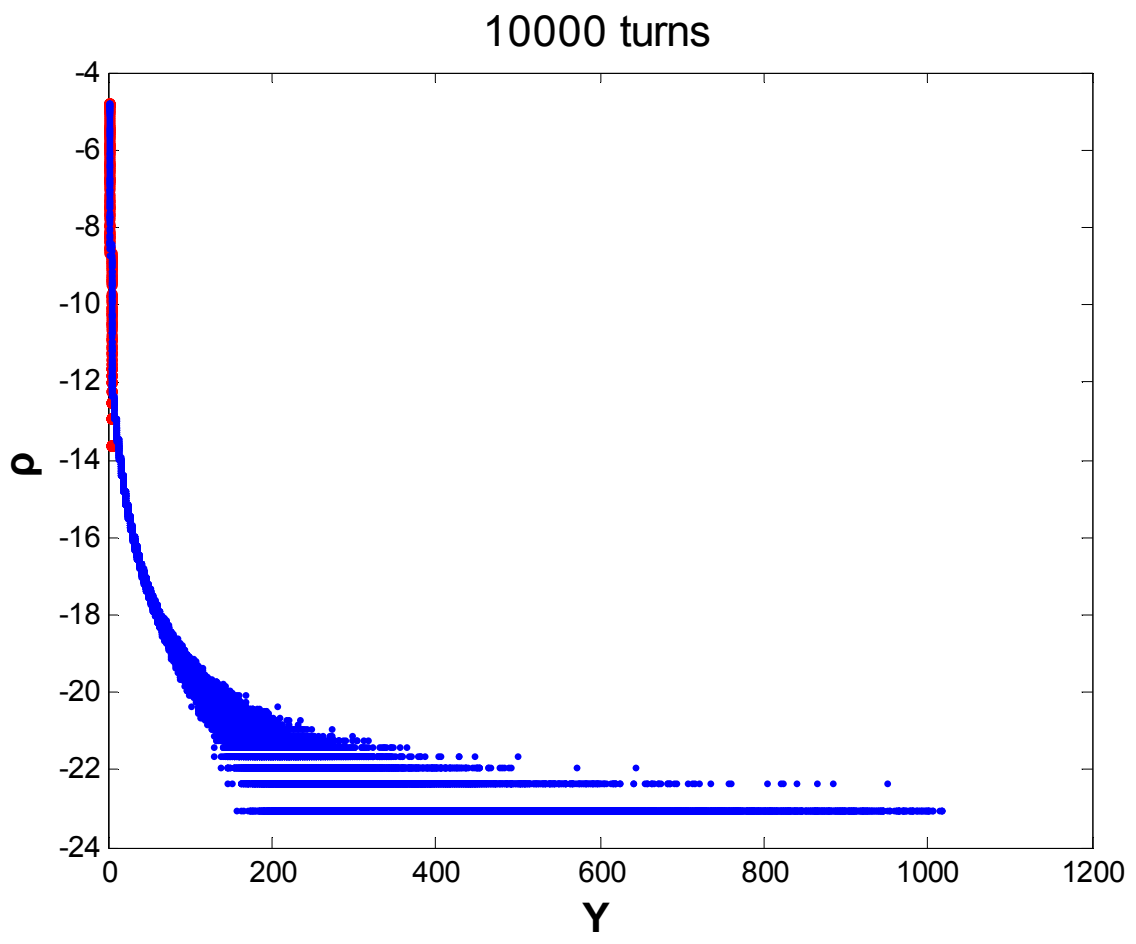

(b)

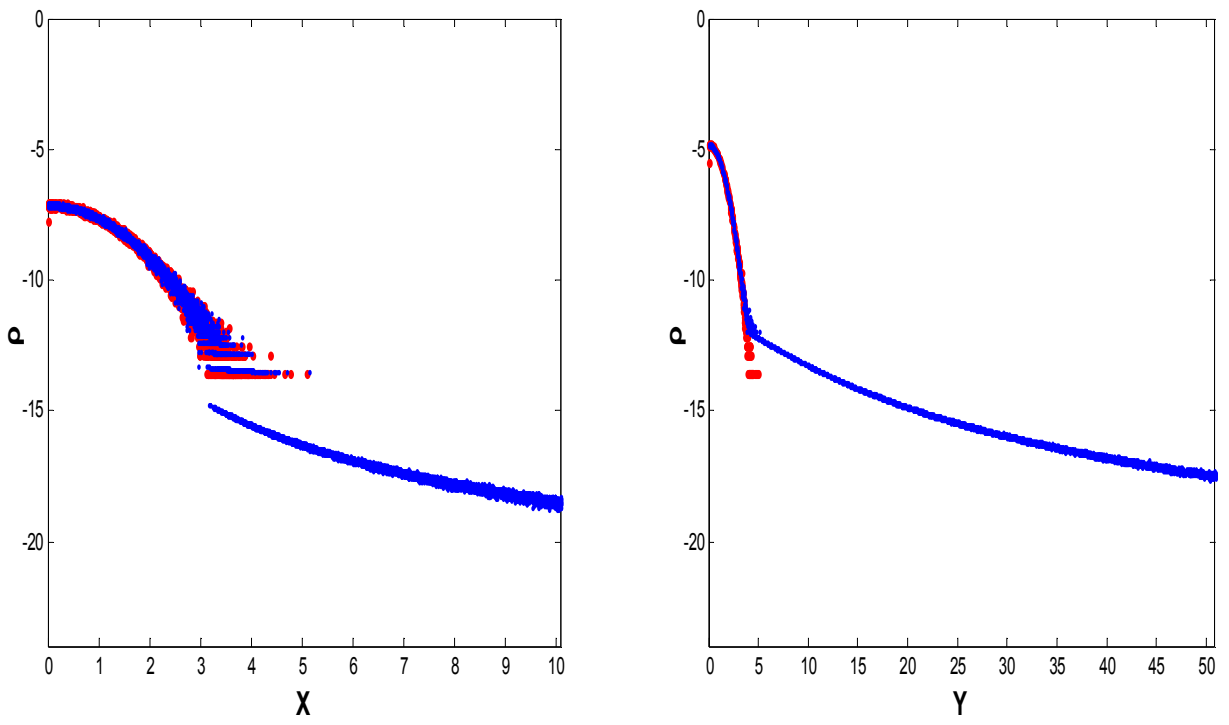

(c)

Fig.9 the statistical distribution in transverse direction, the horizontal axis X/Y is the distance normalized by the nominal horizontal/vertical beam size. The vertical axis represents the distributions in X/Y using a logarithmic scale. The two pictures in (c) show the details in small distance.

## 5 Summary and outlook

In this paper, by deducing the collision formulas in laboratory coordinate system and choosing appropriate cross section and scattering angle, we simulate the effect of the BGCS in the ultra-low emittance electron storage ring using the PIC-MCC method. The results show that the BGCS not only relates to the beam lifetime but also can cause the transverse diffusion and emittance growth.

In the future work, more random process will be added to the simulation, like more kinds of residual gases, inelastic scattering with residual gas, IBS, synchrotron radiation, etc. On the other hand, we are going to realize program parallelization to improve computational efficiency.

## References

1 Z.P Liu. The physical introduction to Synchrotron radiation. frist edition. Hefei, Anhui Province: USTC Publishing Company, Inc., 2009. 143-157

2 Henry J.Halama. J. Vac. Sci. Technol, 1985, **A3**: 1699

3 http://d.g.wanfangdata.com.cn/Conference_26363.aspx

4 EUN-SAN KIM. Particle Accelerators, 1997. Vol.56, 249-269

5 C.K Birdsall. IEEE TRANSACTIONS ON PLASMA SCIENCE. APRIL 1991, VOL.19, NO.2

6 D. Tskhakaya, K. Matyash. Plasma phys. 2007, 47, No.8-9, 563-594

7 W.G.Vincenti and G.H.Kruger Jr, Introduction to Physical Gas Dynamics. original edition. Malabar, Florida: Krieger Publishing Company, Inc., 1967. 348-353

8 LuXin Yu. The beam gas coulomb scattering in electron storage ring. In: Proceeding of IPAC2013, Shanghai, 2013. 1778-1780

9 L.Reimer. Transmission Electron Microscopy,.. fifith edition. Spring-Verlag, 1984. 141-152

10 Petr.Chaloupka, Zdenek Kluiber. Electron beam.10th IYPT.http://ilyam.org/Kluiber_et_al_10th_IYPT_GM_HK_1998.pdf

11 Spear3 Conceptual Design Report, 2002. 3-90 http://www.slac.stanford.edu/cgi-wrap/getdoc/slac-r-609a.pdf